\documentclass[usenatbib]{basi}
\usepackage[T1]{fontenc}
\usepackage[british]{babel}
\usepackage[varg]{txfonts}
\usepackage{rotating}
\usepackage{dcolumn}

\usepackage{epsfig}

\def\etal{{\it et al.}\thinspace}

\newcommand{\bei}{\begin{itemize}}
\newcommand{\eei}{\end{itemize}}
\newcommand{\bef}{\begin{figure}}
\newcommand{\eef}{\end{figure}}
\newcommand{\ben}{\begin{enumerate}}
\newcommand{\een}{\end{enumerate}}

\newcommand{\beq}{\begin{equation}}
\newcommand{\eeq}{\end{equation}}
\newcommand{\ber}{\begin{eqnarray}}
\newcommand{\eer}{\end{eqnarray}}

\begin{document}

\title[Magnetic Fields of Neutron Stars]{Glitch Statistics of Radio Pulsars : Multiple populations}

\author[Konar \& Arjunwadkar]%
       {Sushan Konar$^1$ and Mihir Arjunwadkar$^2$\thanks{email: $^1$sushan@ncra.tifr.res.in, $^2$mihir@cms.unipune.ac.in} \\
        $^1$NCRA-TIFR, University of Pune Campus, Pune, India \\
        $^2$Centre for Modeling and Simulation, University of Pune, Pune, India}

\pubyear{2014}
\volume{**}
\pagerange{**--**}
%\pagerange{\pageref{firstpage}--\pageref{lastpage}}
%\status{submitted}
%\date{Received --- ; accepted ---}
\maketitle

\label{firstpage}

\vspace{-1em}
\begin{abstract}
We present statistical evidence suggesting more than one population in
the energy distributing of  pulsar glitches, which implies the presence
of different mechanisms  accessing different energy ranges responsible
for glitches.
\end{abstract}

A glitch is a timing irregularity of radio pulsars, marked by a sudden
increase in the  spin-frequency $\nu$, often followed  by a relaxation
towards the unperturbed  $\nu$.  So far, a total of  451 glitches have
been    seen    in    158    objects    (151    radio    pulsars,    7
magnetars)~\citep{espin11}.   These are  likely caused  by sudden  and
irregular  transfer of  angular momentum  to  the solid  crust of  the
neutron star  by a  super-fluid component rotating  faster; or  by the
crust quakes.  It  is conjectured that the bimodality  seen within the
range of glitch values ($10^{-12} \leq \delta \nu / \nu \leq 10^{-4}$)
are indicative  of these  two separate mechanisms~\citep{yu13}.   As a
step  towards  understanding  the mechanism  underlying  glitches,  we
consider the  statistical nature  of the  glitch energy  ($E_{\rm g}$)
distribution.   The rotational  energy ($E_{\rm  r}$) of  a pulsar  is
approximately $I \nu^2$,  where $I$ is the stellar  moment of inertia.
The change in rotational  energy due to a glitch is  then $E_{\rm g} =
\delta E_{\rm r}  \simeq I \, \nu  \, \delta \nu$, assuming  $I$ to be
roughly  constant ($\simeq  10^{45}$~gm.cm$^2$)  across the  glitching
pulsar population.

One possible way in which  multiple glitching populations may manifest
in the data is via  multimodal structure in the probability densities.
We therefore apply three standard statistical tests to the $\log_{10}(
\delta \nu / \nu )$ and  $\log_{10}( \delta E)$ data; namely, dip test
\citep{HH1985}, Silverman  test \citep{Silverman1981},  and bimodality
test \citep{HV2008}. The  null hypothesis for all these  tests is that
of unimodality.  As a measure of evidence against the null hypothesis,
we report the $p$-value for each test ($0 \le p \le 1$; lower the $p$,
greater  the   evidence  against   the  null  hypothesis).    For  the
$\log_{10}(  \delta \nu  / \nu  )$ data,  the $p$-values  are $\approx
10^{-3},  0,$ and  $0$  % 7  \times  10^{-4} respectively,  suggesting
strong evidence against unimodality.  For  the $\log_{10}( \delta E )$
data,   the  $p$-values   are  $\approx   0.14,  0,$   and  $10^{-10}$
respectively.  The dip  test $p$-value, though closer to 0  than to 1,
suggests little or no evidence  against unimodality.  Reasons for this
(apparent) discrepancy  between the dip  test and the other  two tests
are being investigated.

On  the  other  hand,  unimodal structure  of  a  probability  density
function  may  not  necessarily  rule out  the  presence  of  multiple
populations.   For  example,  the  sum  of  two  not-so-well-separated
Gaussians  may  be  unimodal.   We  therefore  model  the  probability
densities  of  the two  glitch  quantities  as mixtures  of  Gaussians
\citep{MP2000} with  up to 5  components
(Gaussianity is a purely modeling assumption here).
We apply  a multimodel bootstrap  approach coupled
with  BIC-based   model  selection  \citep{BA2002}  to   obtain  model
selection frequencies.  Here, selection  frequency for the 1-component
model is  akin to the  $p$-value for a test.   The two columns  on the
left   in  Fig.\   \ref{stat}  show   data  histograms   with  mixture
fits. BIC-optimal  mixture size is  2 for both quantities.   The right
column  shows   model  selection  frequencies  against   mixture  size
(bootstrap  size: 10000):  We see  that the  1-component model  is the
least favoured one for either quantity.

\bef
 \centerline{\includegraphics[angle=-90,width=0.865\textwidth]{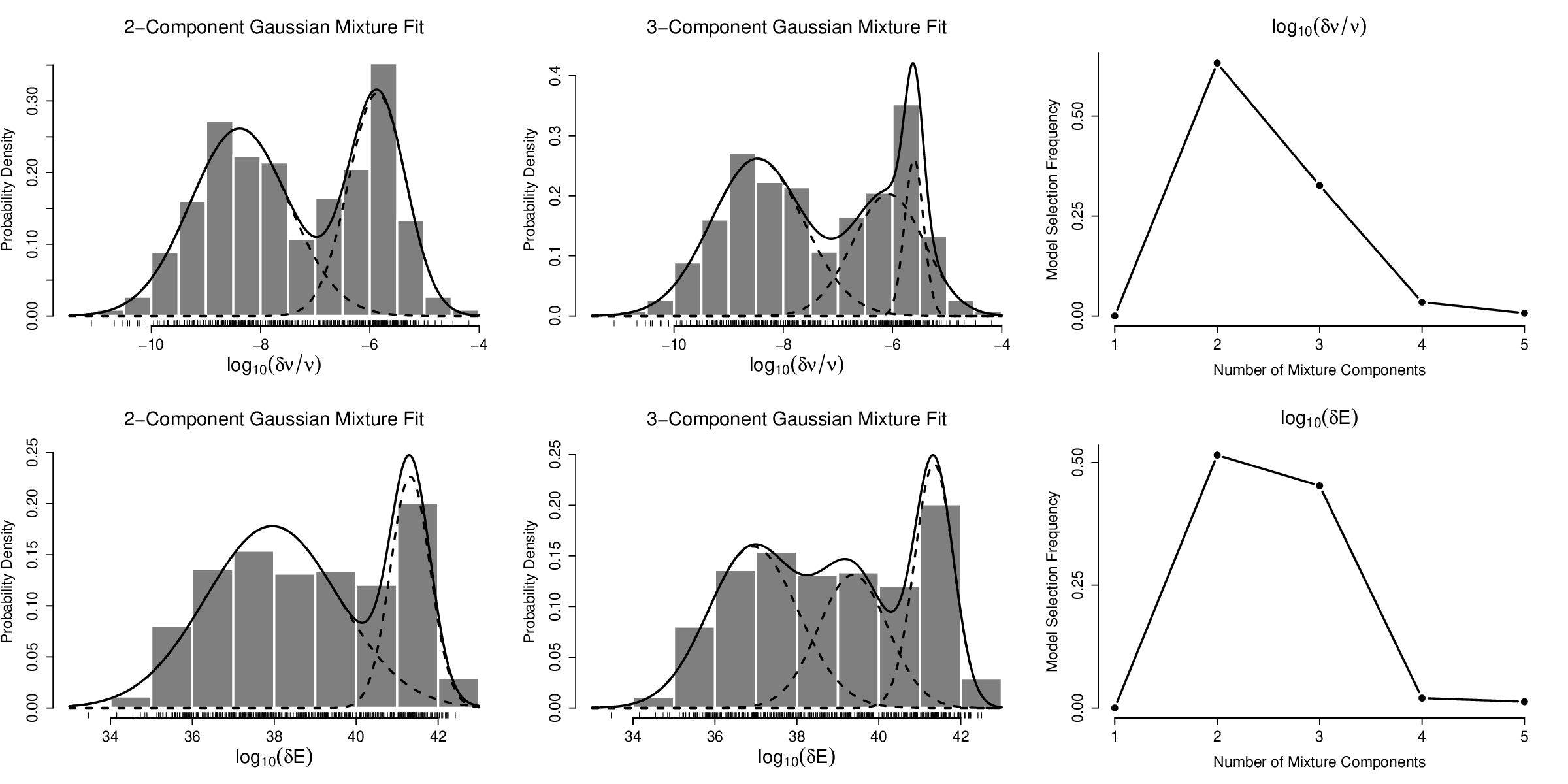}} % latex, dvips
 % \centerline{\includegraphics[width=0.865\textwidth]{gltc.pdf}}         % pdflatex
%
\caption{\label{stat}Statistical  evidence  suggesting more  than  one
  population; see text for description.}
\eef

Our statistical  analysis is therefore  indicative of the  presence of
structure in the  data which, in turn, is suggestive  of more than one
glitch mechanisms corresponding to different intrinsic energies. These
may well  correspond to  the energies  available in  different crustal
regions  of   a  neutron   star~\citep{manda09}.   We   conclude  that
mechanisms  responsible   for  glitches  are  perhaps   different  for
different  energy regimes,  originating  in different  regions of  the
star.  (We used the \texttt{R} statistical
computing environment \citep{R2013} for computation.)

%
%Data used in this paper is available at {\scriptsize \texttt{http://www.ncra.tifr.res.in:8081/$\sim$sushan/nsws.html}}.)

%

\end{document}